\documentclass[twocolumn,showpacs,amsmath,amssymb,prl,nobibnotes,floatfix
]{revtex4-1}

\usepackage{bm}
\usepackage[plainpages=false]{hyperref}
\usepackage{graphicx}
\usepackage{multirow}

\usepackage[usenames,dvipsnames]{color}
\usepackage[normalem]{ulem}

\usepackage{dcolumn}
\newcolumntype{d}[1]{D{.}{.}{#1}}

\def\fun#1#2{\lower3.6pt\vbox{\baselineskip0pt\lineskip.9pt
  \ialign{$\mathsurround=0pt#1\hfil##\hfil$\crcr#2\crcr\sim\crcr}}}

 \newcommand{\mnras}{Mon.\ Not.\ R.\ Astron.\ Soc.\ }
 
 \newcommand{\jcap}{J.\ Cosmol.\ Astropart.\ Phys.\ }
 \newcommand{\physrep}{Phys.\ Rep.\ }
 \newcommand{\araa}{Annual.\ Rev.\ Astron.\ Astrophys.\ }
 \newcommand{\apjs}{Astrophys.\ J.\ S.\ }

\begin{document}

\title{Axion-like Dark Matter Constraints from CMB Birefringence}

\author{G\"unter Sigl$^{1}$}
\email[]{guenter.sigl@desy.de}

\author{Pranjal Trivedi$^{1,2,3}$}
\email{pranjal.trivedi@desy.de}

\affiliation{$^{1}$Universit\"at Hamburg, {II}. Institut f\"ur Theoretische Physik, Luruper Chaussee 149, 22761 Hamburg, Germany.\\
$^{2}$Hamburger Sternwarte, Gojenbergsweg 112, 21029 Hamburg, Germany.\\
$^{3}$Department of Physics, Sri Venkateswara College, University of Delhi 110020 India}

\begin{abstract}
Axion-like particles are dark matter candidates motivated by the Peccei-Quinn mechanism and also occur in effective field theories where their masses and photon couplings are independent. 
We estimate the dispersion of circularly polarized photons in a background of oscillating axion-like particles (ALPs) with the standard $g_{a\gamma}\,a\,F_{\mu\nu}\tilde F^{\mu\nu}/4$ coupling to photons. 
This leads to birefringence or rotation of linear polarization by ALP dark matter.  
Cosmic microwave background (CMB) birefringence limits $\Delta \alpha \lesssim (1.0)^\circ$ enable us to constrain the axion-photon coupling $g_{a\gamma} \lesssim 10^{-17}-10^{-12}\,{\rm GeV}^{-1}$, for ultra-light ALP masses $m_a \sim 10^{-27} - 10^{-24}$ eV. 
This improves upon previous axion-photon coupling limits by up to four orders of magnitude. 
Future CMB observations could tighten limits by another one to two orders.
\end{abstract}


\maketitle

\textit{Introduction}: 
Axion-like particles (ALPs) are generally understood as pseudoscalar fields $a$ with a two-photon coupling of the form $g_{a\gamma}\,a\,F_{\mu\nu}\tilde F^{\mu\nu}/4$, among other possible couplings that are typically less relevant in dilute media. 
ALPs are generalizations of axions which were originally motivated by solving the strong CP problem through promoting the CP-violating phase $\theta$ to $a/f_a$ with $f_a$ known as the Peccei-Quinn scale \cite{Peccei:1977PhRvL..38.1440P,Weinberg:1978PhRvL..40..223W,Wilczek:1978PhRvL..40..279W}. 
Through its couplings to gluons and quarks the axion will attain a mass $m_a$ below the QCD scale through which its expectation value is driven to zero. ALPs also arise naturally in low energy effective field theories of string compactifications. \cite{Svrcek:2006yi,Conlon:2006tq,Arvanitaki10:PhysRevD.81.123530,Cicoli:2012sz}.

Apart from the coupling constant $g_{a\gamma}$, ALPs are characterized through their vacuum mass $m_a$. 
In contrast to axions, ALPs, in general, do not solve the strong CP problem and $g_{a\gamma}$ and $m_a$ are assumed to be independent parameters.
In addition to the misalignment mechanism generating ALP cold dark matter \cite{Preskill:1982cy,Abbott:1982af,Dine:1982ah,VisinelliGondolo2009:PhysRevD.80.035024,Marsh:2015xka}, inflation also produces ALP field fluctuations of order the inflationary expansion rate, $H_i/(2\pi)$, which gives a contribution $\rho_a \sim m_a^2 H_i^2/(2\pi)^2$. For $m_a \gtrsim 3\times 10^3 (10^{16} {\rm GeV}/H_i) \,H(z_{\rm rec})$ and sufficiently large $H_i$, it can also lead to the correct relic dark matter density.

The coupling term can lead to ALP-photon oscillations in the presence of external electromagnetic fields and also lead to an effective refractive index for photons propagating in a background of ALPs. 
While the former effect has been investigated extensively in both cosmological, astrophysical contexts \cite{JaeckelRingwald2010:2010ni,Arias:2012az,Marsh:2015xka} and in experiments \cite{Graham:2015ouw,Irastorza:2018dyq}, at least so far, the latter effect is less well studied. 
This refractive effect can be particularly relevant if ALPs constitute a part of dark matter which is what we assume without specifying the ALP production processes.

We investigate the birefringent effect of ALP dark matter on the cosmic microwave background (CMB) which is well constrained observationally. 
The linear polarization of the CMB is sourced by the quadrupole temperature anisotropy of the radiation field via Thomson scattering \cite{Hu2002ARA&A..40..171H}. 
The curl-free E-mode polarization \cite{Kovac:2002Natur.420..772K} has been observed by Planck \cite{PlanckParameters2018arXiv180706209P} and several ground-based experiments at higher resolution \cite{Staggs:2018RPPh...81d4901S}. 
Steadily improving upper limits have also been placed on the gradient-free B-mode polarization in the quest for the signal of inflationary tensor perturbations \cite{Kamionkowski18ARAA}. 
The birefringence or rotation of linear polarization arising due to any parity violation \cite{Carroll1990:PhysRevD.41.1231,Carroll:1991zs,Harari:1992ea,Carroll:1998PhysRevLett.81.3067,Galaverni:2014gca}, in the case of the CMB  \cite{LueWangKamionkowski:1999PhRvL..83.1506L,LiuLeeNg2006:PhysRevLett.97.161303,Feng2006:PhysRevLett.96.221302,Finelli2009PhRvD..79f3002F,Giovannini2005:PhysRevD.71.021301}, can be probed via correlations of CMB E and B-modes \cite{Kamionkowski:2009PhRvL.102k1302K,Finelli2009PhRvD..79f3002F,Yadav:2012PhRvD..86h3002Y}. 

In this letter we calculate the overall amplitude of birefringence predicted by an oscillating background of ALP dark matter and use the observational limits on CMB birefringence to constrain the axion-photon coupling constant as a function of ALP mass. 
This leads to significantly improved constraints on $g_{a\gamma}$ in the mass range $10^{-27}\,{\rm eV}\lesssim m_a\lesssim10^{-23}\,{\rm eV}$, overlapping with the mass range of ultra-light and fuzzy dark matter \cite{Hu2000FuzzyCDM:PhysRevLett.85.1158,Hui:2017PhRvD..95d3541H} which has gained significant attention recently. 
The prior constraints are generally flat in $m_a$ for $m_a \lesssim 10^{-10}$ eV and at the level $\sim 10^{-12}$ GeV$^{-1}$ \cite{Berg:2017ApJ...847..101B,MarshMCD:2017JCAP...12..036M}.
Here, we ignore clustering of ALP dark matter \cite{Marsh:2015xka,SakharovKhlopov:1994id,Enander:2017ogx,Vaquero:2018tib,Veltmaat:2018dfz} and the anisotropies in the birefringence \cite{LiZhang08:2008tma,Gluscevic2012:PhysRevD.86.103529,POLAREBAR2015PhRvD..92l3509A,Contreras2017:2017JCAP...12..046C,Liu2017:2017PDU....16...22L}.

\textit{Photon Propagation in an ALP background}:  
\label{sec:matheiu}
Using natural units $c=\hbar=k_B=1$ and Lorentz-Heaviside units $\epsilon_0=\mu_0=1$,
the parts of the Lagrangian depending on the ALP and photon fields can be written as
\begin{equation}\label{eq:L_a}
  {\cal L}_{a\gamma}=-\frac{1}{4}F_{\mu\nu}F^{\mu\nu}+\frac{1}{2}\partial_\mu a\partial^\mu a+
  \frac{1}{4}g_{a\gamma}\,a\,F_{\mu\nu}\tilde F^{\mu\nu}-V_a(a)\,,
\end{equation}
where $F_{\mu\nu}$ is the electromagnetic field strength tensor, $\tilde F_{\mu\nu}$ is its dual and $V_a(a)$ is the
effective ALP potential which can be expanded as $V_a(a)=\frac{1}{2}m_a^2a^2+{\cal O}(a^3)$ around $a=0$,
with $m_a$ the effective ALP mass. For the axion-photon coupling, one often uses the relation
\begin{equation}\label{eq:f_a}
  g_{a\gamma}=\frac{s\alpha_{\rm em}}{2\pi f_a}\,,
\end{equation}
with $s$ as a model dependent parameter of order unity, $\alpha_{\rm em}$ the fine structure constant
and $f_a$ the Peccei-Quinn scale.
For left- and right-circular polarization photon modes propagating in the $z-$direction, we make the Ansatz
${\bf A}_\pm(t,{\bf r})=A_\pm(t){\bf e}_\pm e^{ikz}\,,$
where ${\bf e}_\pm\equiv{\bf e}_x\pm i{\bf e}_y$ are the unit vectors corresponding to left and right-circular modes.
To zeroth order, photon wave-packets propagate along trajectories $z=t$ so we can
identify time scales with length scales. Our Ansatz yields the equation of motion for the photon modes in an ALP background. We can derive the general dispersion relation
\begin{equation}\label{eq:dispersion}
    \omega \simeq k \mp \frac{g_{a\gamma}}{2} m_a \,a_0,
\end{equation}
where 
$a_0$ is the amplitude of the ALP field which is supposed to vary on time and lengths scales much larger than $1/k$ and the inverse photon frequency. 

\textit{Birefringence phase shift and axion-photon coupling}: 
For the ALP amplitude $a$ we use the relation for ALP energy density $\rho_a= (1/2)\,m_a^2\,a^2\,$
and $\rho_a(z_{*}) \simeq F \, \Omega_c \, (1+z_*)^3 \, \rho_{\rm crit,0} $. 
Here, $F = \Omega_a / \Omega_c$ is the ALP dark matter energy density fraction with $\Omega_i = \rho_{i,0}/\rho_{\rm crit,0}$ the fractional energy density at the present epoch and $z_*$ is the redshift of last scattering.
From Eq.~(\ref{eq:dispersion}) we can write the phase difference between the left and right-circularly polarized photon modes as
\begin{equation}\label{eq:Delta_a}
    \Delta \phi=\Delta\omega dt \sim g_{a\gamma} \Delta a
\end{equation}
which is the same as, e.g., Eq.~(2) in \cite{Pospelov2009PhRvL.103e1302P} noting that our $g_{a\gamma}$ is their $1/f_a$ without the factors in our Eq.(\ref{eq:f_a}).
We take $\Delta a\ = \left[a(z_*)-a_{\rm local}\right]$
i.e. the difference of the axion field values at the surface of last scattering and locally \cite{Carroll1990:PhysRevD.41.1231,Carroll:1991zs,Carroll:1998PhysRevLett.81.3067,Harari:1992ea,Arvanitaki10:PhysRevD.81.123530,Fedderke:2019ajk} and can thus constrain the absolute value of $g_{a\gamma}$. 

In the supplementary material \cite{SM} we clarify whether a random walk behaviour of the birefringence phase angle will occur for photon propagation over multiple ALP coherence lengths (cf. Refs.~\cite{Fedderke:2019ajk,Caputo:2018ljp,Fujita:2018arXiv181103525F,Liu:2019brz,Ivanov:2018byi}). 
Although not applicable here, we show that a random walk is possible, in principle, from non-linear terms in the solution to the modified photon wave equation. 
These terms are suppressed by a large factor $(m_a/k)$ compared to the linear term producing just the difference in ALP field values, for the very low ALP masses and CMB frequencies considered here. 
We note in passing that a random walk in angle might also be relevant if there are discontinuities in the ALP field gradient (from domains or cosmic strings) or if there are multiple ALPs coupled to electromagnetism.

We note that $\Delta\phi$ does not depend on the photon frequency (unlike Faraday rotation) so that observations at any frequency can be applied. 
Using the local Galactic dark matter density $\rho_{a,\,{\rm local}} \simeq 0.3 \,{\rm GeV}\,{\rm cm}^{-3}$ and cosmological parameter values from Planck \cite{PlanckParameters2018arXiv180706209P}, we note that 
$a(z_*)/a_{\rm local} \sim 74$ and after subtracting the local ALP field value, we get
\begin{eqnarray}\label{eq:a_z_*}
   \Delta a &\simeq& 
   5.22 \times 10^{15} 
\!\left( \frac{m_{a,\,{\rm ref}}}{m_a} \right)
   F^{1/2}
\!\!\left( \frac{\Omega_c}{0.264} \right)^{\!1/2} \nonumber \\
&\times& \left(\frac{\rho_{\rm crit,0}}{8.098\times 10^{-47} h^2 \,{\rm GeV}^4}\right)^{1/2}
\!\left(\! \frac{1+z_*}{1090} \!\right)^{\!\!3/2} \!\!{\rm GeV}\!,  
\end{eqnarray}
evaluated at a reference ALP mass $m_{a,\,{\rm ref}} = 3 \times 10^{-26}$ eV.

The observed birefringence angle $\Delta\alpha$ (the angle of rotation of linear polarization) is half this phase shift  $\Delta \phi$ between the two components of circular polarization. 
Observed and forecast limits on $\Delta\alpha$ from CMB observations are listed in Table \ref{tab:CMB}. 
The rotation of polarization (as well as its angular power spectrum) can be extracted via measuring correlations between E and B-modes \cite{Kamionkowski:2009PhRvL.102k1302K} which arise when there is parity violation, in our case sourced by the pseudoscalar ALP field. 
To derive constraints on the axion-photon coupling, we adopt a conservative limit $\Delta\alpha \lesssim (1.0)^\circ$, between the current Planck and SPTpol limits and the ACTpol, BICEP2/Keck Array and POLARBEAR limits. We add the absolute value of the quoted rotation angle and its errors, for each experiment, to estimate the limits on $g_{a\gamma}$ in Table~\ref{tab:CMB}. 

The birefringence limit leads to an upper limit constraint on the axion-photon coupling, 
\begin{eqnarray}\label{eq:g_constr}
g_{a\gamma}& \lesssim&
1.60 \!\times\!10^{-15}
\!\left( \!\frac{\Delta\alpha}{\,1.0^{^{\circ}}}\!\right)
\!\left( \frac{m_a}{m_{a,{\rm ref}}} \right)
\!\!\left( \frac{10^{-2}}{F} \right)^{\!1/2} 
\!\!\left( \frac{0.264}{\Omega_c} \right)^{\!1/2} \nonumber \\
&\times&
\left( \! \frac{0.674}{h} \!\right)
\left(\! \frac{1090}{1+z_{\rm rec}} \!\right)^{\!\!3/2} 
\left(\! \frac{R\,(m_a)}{R\,(m_{a,{\rm ref}})} \!\right){\rm GeV}^{-1}.
\end{eqnarray}
From the limits on $F$ derived from the CMB power spectrum  \cite{Amendola:2005ad,Hlozek:2015,Hlozek:2018}, we adopt conservative values of $F$ as $10^{-2}$ transitioning to $10^{-1}$ over our mass range $7 \times 10^{-28} - 3 \times 10^{-24}$ eV. 
Our constraints on $g_{a\gamma}$ depend relatively weakly on ALP dark matter fraction $F$ limits.

The ALP field oscillates in time at a period $\tau_{osc} = 2\pi/m_a \simeq (1.31 \times 10^{-27}$ eV/ $m_a$)($10^5$ yr), over the extended recombination epoch lasting $\tau_{rec} \sim 10^5$ yr. 
As a result, the birefringence rotation angle sourced by the ALP field is averaged over its oscillating phase \cite{Fedderke:2019ajk}. 
The net polarization rotation angle that survives is proportional to the uncancelled fraction of the ALP oscillations during recombination.
This leads to a power-law reduction of rotation angle, described as a washout effect by Ref.~\cite{Fedderke:2019ajk}. 
\begin{widetext}

\begin{table}
 \centering
\begin{tabular}{lccccc}
 \toprule
 CMB Exp. & Frequency (GHz) \,\, & \,\, \,\,\,$\ell$ \,\,\,\,\,\,& $\Delta\alpha$ (degrees) & \,\,$\Delta\alpha$ Reference & \, $(g_{a\gamma} 10^{15})$ GeV$^{-1}$  \\
 \toprule
 WMAP7 & 41,61,94 & 2-800 & $-1.1 \pm1.4\,(\pm 1.5)$ & Komatsu et al.(2011)  \cite{Komatsu:2011ApJS..192...18K} & 6.3\\ 
 BICEP2/Keck & 150 & 30-300 & $1.0 \,(\pm0.2)$  & BICEP2 (2014)  \cite{BICEP14PRL_Ade:2014xna} & 1.9 \\
 ACTpol & 146 & 250-3025 &  $0.29\pm0.28\,(\pm0.5)$ & Molinari et al. (2016)  \cite{Molinari2016PDU....14...65M} & 1.7 \\
 Planck & 30-857 (9 ch.) & 2-1500 & $0.31 \pm0.05\,(\pm0.28)$ & Planck (2016) XLIX  {\cite{Planck16ParityViolation_Aghanim:2016fhp}} & 1.0 \\
 POLARBEAR & 148 & \,\,\,500-2700\,\,\, & -0.73 $\pm0.17\,(\pm0.56)$ & POLARBEAR (2017)  \cite{POLARBEAR17_Ade:2017uvt} & 2.3 \\
 SPTpol & 150 & 100-2000 & 0.63 $(\pm0.04)$ & Wu et al. (2019)  \cite{SPTpol19_Wu:2019hek} & 1.1 \\
 {\it Current status} &  &  & 1.0 & {\it Limit adopted for Eq.~(\ref{eq:g_constr})}  & 1.6 \\
(CMB-S4) & 35-250 & $\lesssim10^4$  &  $0.03$ & CMB-S4 Science Book (2016) \cite{CMBS4:2016arXiv161002743A} & 0.048 \\
 (COrE) & 135 (60-600)& $\lesssim$1400 & $0.003$ & Molinari et al. (2016)  \cite{Molinari2016PDU....14...65M}  &  $4.8 \cdot 10^{-3}$\\
  (PICO) & 70-156 (21-799)& $\lesssim$1400 & $0.0017$ & Hanany et al. (2019)  \cite{PICO19_Hanany:2019lle}  &  $2.7 \times 10^{-3}$ \\
 (CV-lim. r=0.1)&  & $\lesssim$3000 &  $10^{-5}$ & Molinari et al. (2016)  \cite{Molinari2016PDU....14...65M}  & $1.6 \times 10^{-5}$ \\
 \toprule
\end{tabular}
\caption{Comparison of observed limits and forecasts on birefringence angle $\Delta \alpha \pm \text{ stat. error} \,\,(\pm \text{ syst. error})$ from various CMB experiments (and a cosmic variance-limited case for tensor-to-scalar ratio $r$=0.1) with their frequency and multipole $\ell$ range. 
We add the absolute values of the quoted rotation angle and its errors, for each experiment, to estimate the corresponding limits on axion-photon coupling $g_{a\gamma}$, derived using Eq.~(\ref{eq:g_constr}), evaluated at $m_a = m_{a,{\rm ref}} \simeq 3 \times 10^{-26}$ eV.}
\label{tab:CMB}
\end{table}

\end{widetext}
We model this effect as a polarization rotation angle reduction factor $R\,(m_a) = 1 + (\tau_{rec}/\tau_{osc}) \simeq 1 + (m_a / 1.31 \times 10^{-27}$ eV) $\simeq 1.5-2000$, over our range of very small ALP masses where $\tau_{osc}$ can be a bit larger to significantly smaller than $\tau_{rec}$. 
Due to $R\,(m_a)$, the coupling now scales as $g_{a\gamma} \propto m_a^2$ for $m_a \gg 10^{-27}$ eV. 
However, even at the high mass end with the largest $R\,(m_a)$, the weakest portion of  birefringence constraints derived for $g_{a\gamma}$, after washout, are still comparable to the tightest previous constraints $\sim 10^{-12}$ GeV$^{-1}$.

We only consider ALP masses above the temperature-independent limit for ALP dark matter \cite{RPP2018Tanabashi} is $m_a \gtrsim 7 \times 10^{-28}$ eV. 
We ignore smaller masses as at $m_a \lesssim H(z_{\rm rec}) \sim 3\times 10^{-29}$ eV, the ALP would not oscillate at recombination, acting as dark energy till $H(z <z_{\rm rec}) \lesssim m_a$.

\begin{widetext}

\begin{figure}
\centering
\includegraphics[width=0.99\columnwidth]{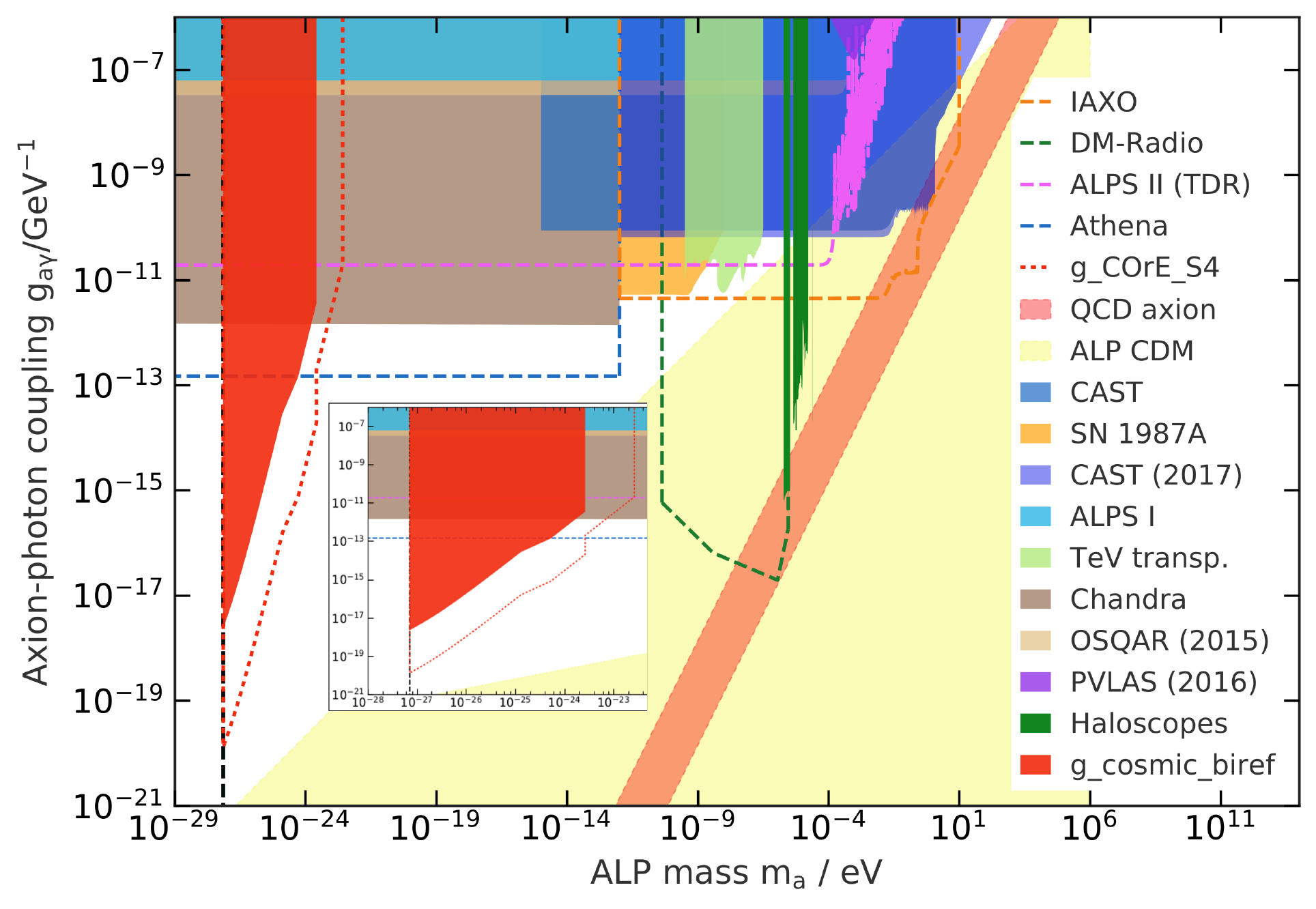}
\caption{Axion-photon coupling constant $g_{a\gamma}$ as a function of ALP dark matter mass. 
The CMB birefringence cosmological constraints on $g_{a\gamma}$ from Eq.~(\ref{eq:g_constr}) (filled red, labelled g\_cosmic\_biref, also expanded in the inset figure) improve substantially on x-ray bounds from Chandra \cite{Berg:2017ApJ...847..101B} (filled brown) and forecasts for Athena \cite{Conlon:2018MNRAS.473.4932C} (blue dashed line). 
Future CMB observations from COrE and CMB-S4 (dotted red line) could further improve these $g_{a\gamma}$ constraints by a factor $\sim$ 340 or 34, respectively.
Other filled regions (or dashed lines) depict parameter space already excluded (or future forecasts) due to various experiments and observations \cite{Alplot} (some based on ALP-photon oscillations, independent of ALP dark matter). 
The vertical black dashed line (at 7 $\times 10^{-28}$ eV) indicates the temperature-independent limit of ALP dark matter \cite{RPP2018Tanabashi}. 
The yellow filled region is for canonical temperature-dependent ALP dark matter via the misalignment mechanism \cite{Ringwald:2013via}. The light red band with unit slope depicts QCD axion models \cite{Kim:1979if,Shifman:1979if,Dine:1981rt,Zhitnitsky:1980tq} (plot created with the ALPlot code \cite{Alplot}).}
\label{fig:g_m_plot}
\end{figure}

\end{widetext}

The multipole $\ell$ of the signal is $\ell \sim \pi/\Theta \sim \pi d_A/l_c \sim \pi d_A m_a v_a(z_{\rm rec}) \sim 2\pi m_a v_a(z_{\rm rec})/[H_0 (1+z_{\rm rec})]$, where $l_c$ is the ALP coherence length, $v_a$ the ALP velocity $\sim 10^{-4}$ \cite{Tseliakhovich:2010PhRvD..82h3520T} and $d_A$ is the angular diameter distance.
Therefore, the highest mass angular scale resolved by a maximum observed multipole $\ell_{\rm max} \sim 10^3$ is given by  $[\ell_{\rm max} H_0 (1+z_{\rm rec})]/ [2\pi v_a(z_{\rm rec})]$ or $m_a \lesssim  2.6 \times 10^{-24}$ eV. 
The $g_{a\gamma}$ vs $m_a$ region excluded over this mass range using Eq.~(\ref{eq:g_constr}) is depicted by the filled red region in Fig.~\ref{fig:g_m_plot} and its inset.

We note that our upper limit $g_{a\gamma} \lesssim 1.6 \times 10^{-15}$ GeV$^{-1}$ at $m_{a,{\rm ref}} \sim 3 \times 10^{-26}$ eV is $\sim 10^3$ times tighter than previous limits from a search for oscillations in x-ray spectra of active galactic nuclei in magnetized clusters (NGC 1275 in the Perseus cluster \cite{Berg:2017ApJ...847..101B} or M 87 \cite{MarshMCD:2017JCAP...12..036M}) and even $\sim 10^2$ times stronger than the limit forecast for Athena \cite{Conlon:2018MNRAS.473.4932C}. 
However, it must be emphasized that these x-ray spectra and CMB birefringence limits depend on wholly different assumptions - the x-ray limits do not assume ALP dark matter and assume a 25 $\mu$G cluster center magnetic field for Primakoff conversion . 
The ALP birefringence effect is independent of any magnetic fields. 
Our constraints are also comparable in magnitude to the constraints inferred from polarized emission from a nearby protoplanetary disk \cite{Fujita:2018arXiv181103525F}. 
We expect that CMB birefringence constraints could be more robust than other methods involving polarized astrophysical sources as they are relatively unaffected by systematic uncertainties in determining the source's intrinsic polarization angle. 

Future CMB observations promise to greatly reduce the r.m.s. uncertainty in the birefringence angle (Table~\ref{tab:CMB}). 
Controlling polarization-angle calibration systematics \cite{Kaufman:2016MNRAS.455.1981K,Miller:2009PhRvD..79f3008M}, currently at the level of $(0.3)^\circ$, will be crucial \cite{Pagano2009PhRvD..80d3522P} to achieving the $\sim$30 to 500 times tighter $\Delta \alpha$ \cite{Molinari2016PDU....14...65M} and $g_{a\gamma}$ limits that are forecast for the CMB-S4, COrE and PICO-like experiment.

We also note that rotation of polarized synchrotron emission from radio galaxies \cite{Carroll:1998PhysRevLett.81.3067,Leahy:1997astro.ph..4285L,Ivanov:2018byi} and scattered UV emission from extragalactic sources \cite{Alighieri:2010ApJ...715...33D} can also be used to constrain the birefringence angle to approximately $1^\circ$. 
Such sources have the disadvantages of having to correct for Faraday rotation, projections effects and differing intrinsic polarization properties. 
However, with a future SKA 2 survey of $\sim10^6$ polarized sources, a much improved overall birefringence angle error of $2 \times 10^{-3}$ degree has been forecast (similar to COrE) for a maximum multipole  $\ell_{\rm max} \sim 700$ \cite{Whittaker:2018MNRAS.474..460W}.

A time variation of the observed polarization signal, due to oscillations of the local axion field, described by \cite{Fedderke:2019ajk}, can put constraints on $g_{a\gamma}$ at slightly higher ALP masses \cite{Fedderke:2019ajk,Liu:2019brz,Ivanov:2018byi,Caputo:2019tms}. However, for our mass range, the time variation of the signal is several decades at its fastest rendering it unimportant.

\textit{Discussion and Conclusions}: \label{sec:Conclusions}
The effective refractive index for photons propagating in an oscillating ALP dark matter background results in birefringence or a rotation in the linear polarization of radiation. 
The photon dispersion relation contains a term oscillating in time with amplitude proportional to $g_{a\gamma}$. 
The birefringence produced is independent of frequency and can be utilized to place constraints on the axion-photon coupling $g_{a\gamma}$.

We derived the ALP birefringence angle for CMB polarization, taking into account ALP dark matter fraction $F$ and reduction in rotation from ALP oscillations during recombination. 
Setting a current limit on birefringence angle of $(1.0)^\circ$ from CMB observations, we constrain the axion-photon coupling $g_{a\gamma} \lesssim 1.6 \times 10^{-15}$ GeV$^{-1}$ at the reference ALP mass scale $m_{a,{\rm ref}}\sim 3 \times 10^{-26}$ eV. The constraint on $g_{a\gamma}$ scales as $m_a^2$ for $m_a \gtrsim m_{a,{\rm ref}}$. At $m_{a,{\rm ref}}$ our constraint improves on bounds from Chandra \cite{Berg:2017ApJ...847..101B,MarshMCD:2017JCAP...12..036M} by $\sim$ three orders of magnitude and the Athena forecast \cite{Conlon:2018MNRAS.473.4932C} by $\sim $ two orders, although x-ray constraints are independent of ALP dark matter.

Our constraints also scale as $g_{a\gamma} \propto \Omega_a^{-1/2}$ or $F^{-1/2}$ and thus depend somewhat weakly on ALP dark matter fraction $F=\Omega_a/\Omega_c$.
Limits have been placed, $F \lesssim 0.2$ for $10^{-23} {\rm eV} \lesssim m_a \lesssim 10^{-21} {\rm eV}$, from non-observation of decrease in density fluctuations at small scales via the Lyman-$\alpha$ forest \cite{Irsic:2017PhRvL.119c1302I,Kobayashi:2017PhRvD..96l3514K}. Note that Lyman-$\alpha$ limits could be affected by uncertainty in the temperature evolution of the inter-galactic medium.
The ALP fraction can be greater, $F \sim 1$ for $10^{-24} {\rm eV} \lesssim m_a \lesssim 10^{-23} {\rm eV}$. 
The gravitational imprint of ALPs in the CMB power spectrum \cite{Amendola:2005ad,Hlozek:2015,Hlozek:2018} limits $F \lesssim 0.1$ for $10^{-25} {\rm eV} \lesssim m_a \lesssim 10^{-24} {\rm eV}$ and $2-3$ times lower for $10^{-27} {\rm eV} \lesssim m_a \lesssim 10^{-25} {\rm eV}$. Therefore, we have used values of $F=10^{-1}$, $10^{-2}$ for these mass ranges respectively.
Note that not all ALP dark matter models can be constrained via the CMB \cite{Folkerts:2013tua,Kawasaki:2014una}.
Recent constraints have also been placed on the dark matter ALP mass for post-inflationary ALPs via isocurvature perturbations \cite{Feix:2019lpo} and on fuzzy dark matter mass using ultra faint dwarf galaxies \cite{Marsh:2018zyw}.

In spite of the limits on $F$, CMB birefringence constraints on $g_{a\gamma}$ are found to reach orders of magnitude tighter than the previous limits \cite{Berg:2017ApJ...847..101B,MarshMCD:2017JCAP...12..036M} and can improve $g_{a\gamma}$ constraints even for an ALP fraction as low as $F\sim 10^{-8}$.

Cosmological birefringence as a probe of ALPs via their effective refractive index is also quite complementary to other axion experiments such as helioscopes, haloscopes and light-shining-through-wall experiments \cite{Irastorza:2018dyq} in the $g_{a\gamma}-m_a$ parameter space (Fig.~\ref{fig:g_m_plot}). 
Recent proposals to measure the same ALP birefringence effect on terrestrial scales with laser interferometers is predicted to be sensitive to the axion-photon coupling at higher ALP masses around $10^{-10} - 10^{-13}$ eV \cite{Obata:2018vvr,Liu:2018arXiv180901656L,DeRocco:2018jwe,Nagano:2019rbw}.

In the future, CMB experiments like CMB-S4, COrE and PICO have the potential to constrain the axion-photon coupling, via the birefringence effect, by an additional $1-2$ orders of magnitude, respectively.

Note Added: During preparation of this paper for submission, we note that Reference \cite{Fujita:2018arXiv181103525F} appeared which develops an analogous idea but in a different context of protoplanetary disks.

\textit{Acknowledgments}: 
This work has been supported by the Deutsche Forschungsgemeinschaft through the Collaborative Research Center SFB 676 ``Particles, Strings and the Early Universe'' and under Germany's Excellence Strategy - EXC 2121 "Quantum Universe" - 390833306. 
We acknowledge useful conversations with Robi Banerjee, Michael Fedderke, Daniel Grin, Matthias Koschnitzke, Jens Niemeyer, Andreas Pargner, Georg Raffelt, Javier Redondo and Andreas Ringwald. We thank the referee for valuable comments.


\begin{thebibliography}{100}

\bibitem{Peccei:1977PhRvL..38.1440P}
R.~D. {Peccei} and H.~R. {Quinn},
\newblock \prl {\bf 38}, 1440 (1977).

\bibitem{Weinberg:1978PhRvL..40..223W}
S.~{Weinberg},
\newblock \prl {\bf 40}, 223 (1978).

\bibitem{Wilczek:1978PhRvL..40..279W}
F.~{Wilczek},
\newblock \prl {\bf 40}, 279 (1978).

\bibitem{Svrcek:2006yi}
P.~Svrcek and E.~Witten,
\newblock JHEP {\bf 06}, 051 (2006), hep-th/0605206.

\bibitem{Conlon:2006tq}
J.~P. Conlon,
\newblock JHEP {\bf 05}, 078 (2006), hep-th/0602233.

\bibitem{Arvanitaki10:PhysRevD.81.123530}
A.~Arvanitaki, S.~Dimopoulos, S.~Dubovsky, N.~Kaloper, and J.~March-Russell,
\newblock Phys. Rev. D {\bf 81}, 123530 (2010).

\bibitem{Cicoli:2012sz}
M.~Cicoli, M.~Goodsell, and A.~Ringwald,
\newblock JHEP {\bf 10}, 146 (2012), 1206.0819.

\bibitem{Preskill:1982cy}
J.~Preskill, M.~B. Wise, and F.~Wilczek,
\newblock Phys. Lett. {\bf B120}, 127 (1983).

\bibitem{Abbott:1982af}
L.~F. Abbott and P.~Sikivie,
\newblock Phys. Lett. {\bf B120}, 133 (1983).

\bibitem{Dine:1982ah}
M.~Dine and W.~Fischler,
\newblock Phys. Lett. {\bf B120}, 137 (1983).

\bibitem{VisinelliGondolo2009:PhysRevD.80.035024}
L.~Visinelli and P.~Gondolo,
\newblock Phys. Rev. D {\bf 80}, 035024 (2009).

\bibitem{Marsh:2015xka}
D.~J.~E. {Marsh},
\newblock \physrep {\bf 643}, 1 (2016), 1510.07633.

\bibitem{JaeckelRingwald2010:2010ni}
J.~Jaeckel and A.~Ringwald,
\newblock Ann. Rev. Nucl. Part. Sci. {\bf 60}, 405 (2010), 1002.0329.

\bibitem{Arias:2012az}
P.~{Arias} {\em et~al.},
\newblock \jcap {\bf 6}, 013 (2012), 1201.5902.

\bibitem{Graham:2015ouw}
P.~W. Graham, I.~G. Irastorza, S.~K. Lamoreaux, A.~Lindner, and K.~A. van
  Bibber,
\newblock Ann. Rev. Nucl. Part. Sci. {\bf 65}, 485 (2015), 1602.00039.

\bibitem{Irastorza:2018dyq}
I.~G. {Irastorza} and J.~{Redondo},
\newblock Progress in Particle and Nuclear Physics {\bf 102}, 89 (2018),
  1801.08127.

\bibitem{Hu2002ARA&A..40..171H}
W.~{Hu} and S.~{Dodelson},
\newblock \araa {\bf 40}, 171 (2002), astro-ph/0110414.

\bibitem{Kovac:2002Natur.420..772K}
J.~M. {Kovac} {\em et~al.},
\newblock \nat {\bf 420}, 772 (2002), astro-ph/0209478.

\bibitem{PlanckParameters2018arXiv180706209P}
{Planck Collaboration} {\em et~al.},
\newblock ArXiv e-prints  (2018), 1807.06209.

\bibitem{Staggs:2018RPPh...81d4901S}
S.~{Staggs}, J.~{Dunkley}, and L.~{Page},
\newblock Reports on Progress in Physics {\bf 81}, 044901 (2018).

\bibitem{Kamionkowski18ARAA}
M.~Kamionkowski and E.~D. Kovetz,
\newblock Annual Review of Astronomy and Astrophysics {\bf 54}, 227 (2016),
  https://doi.org/10.1146/annurev-astro-081915-023433.

\bibitem{Carroll1990:PhysRevD.41.1231}
S.~M. Carroll, G.~B. Field, and R.~Jackiw,
\newblock Phys. Rev. D {\bf 41}, 1231 (1990).

\bibitem{Carroll:1991zs}
S.~M. Carroll and G.~B. Field,
\newblock Phys. Rev. {\bf D43}, 3789 (1991).

\bibitem{Harari:1992ea}
D.~Harari and P.~Sikivie,
\newblock Phys. Lett. {\bf B289}, 67 (1992).

\bibitem{Carroll:1998PhysRevLett.81.3067}
S.~M. Carroll,
\newblock Phys. Rev. Lett. {\bf 81}, 3067 (1998).

\bibitem{Galaverni:2014gca}
M.~Galaverni, G.~Gubitosi, F.~Paci, and F.~Finelli,
\newblock JCAP {\bf 1508}, 031 (2015), 1411.6287.

\bibitem{LueWangKamionkowski:1999PhRvL..83.1506L}
A.~{Lue}, L.~{Wang}, and M.~{Kamionkowski},
\newblock Physical Review Letters {\bf 83}, 1506 (1999), astro-ph/9812088.

\bibitem{LiuLeeNg2006:PhysRevLett.97.161303}
G.-C. Liu, S.~Lee, and K.-W. Ng,
\newblock Phys. Rev. Lett. {\bf 97}, 161303 (2006).

\bibitem{Feng2006:PhysRevLett.96.221302}
B.~Feng, M.~Li, J.-Q. Xia, X.~Chen, and X.~Zhang,
\newblock Phys. Rev. Lett. {\bf 96}, 221302 (2006).

\bibitem{Finelli2009PhRvD..79f3002F}
F.~{Finelli} and M.~{Galaverni},
\newblock \prd {\bf 79}, 063002 (2009), 0802.4210.

\bibitem{Giovannini2005:PhysRevD.71.021301}
M.~Giovannini,
\newblock Phys. Rev. D {\bf 71}, 021301 (2005).

\bibitem{Kamionkowski:2009PhRvL.102k1302K}
M.~{Kamionkowski},
\newblock Physical Review Letters {\bf 102}, 111302 (2009), 0810.1286.

\bibitem{Yadav:2012PhRvD..86h3002Y}
A.~P.~S. {Yadav}, M.~{Shimon}, and B.~G. {Keating},
\newblock \prd {\bf 86}, 083002 (2012), 1207.6640.

\bibitem{Hu2000FuzzyCDM:PhysRevLett.85.1158}
W.~Hu, R.~Barkana, and A.~Gruzinov,
\newblock Phys. Rev. Lett. {\bf 85}, 1158 (2000).

\bibitem{Hui:2017PhRvD..95d3541H}
L.~{Hui}, J.~P. {Ostriker}, S.~{Tremaine}, and E.~{Witten},
\newblock \prd {\bf 95}, 043541 (2017), 1610.08297.

\bibitem{Berg:2017ApJ...847..101B}
M.~{Berg} {\em et~al.},
\newblock \apj {\bf 847}, 101 (2017), 1605.01043.

\bibitem{MarshMCD:2017JCAP...12..036M}
M.~C.~D. {Marsh} {\em et~al.},
\newblock \jcap {\bf 12}, 036 (2017), 1703.07354.

\bibitem{SakharovKhlopov:1994id}
A.~S. Sakharov and M.~{\relax Yu}. Khlopov,
\newblock Phys. Atom. Nucl. {\bf 57}, 485 (1994),
\newblock [Yad. Fiz.57,514(1994)].

\bibitem{Enander:2017ogx}
J.~Enander, A.~Pargner, and T.~Schwetz,
\newblock JCAP {\bf 1712}, 038 (2017), 1708.04466.

\bibitem{Vaquero:2018tib}
A.~Vaquero, J.~Redondo, and J.~Stadler,
\newblock (2018), 1809.09241.

\bibitem{Veltmaat:2018dfz}
J.~Veltmaat, J.~C. Niemeyer, and B.~Schwabe,
\newblock Phys. Rev. {\bf D98}, 043509 (2018), 1804.09647.

\bibitem{LiZhang08:2008tma}
M.~Li and X.~Zhang,
\newblock Phys. Rev. {\bf D78}, 103516 (2008), 0810.0403.

\bibitem{Gluscevic2012:PhysRevD.86.103529}
V.~Gluscevic, D.~Hanson, M.~Kamionkowski, and C.~M. Hirata,
\newblock Phys. Rev. D {\bf 86}, 103529 (2012).

\bibitem{POLAREBAR2015PhRvD..92l3509A}
P.~A.~R. {Ade} {\em et~al.},
\newblock \prd {\bf 92}, 123509 (2015), 1509.02461.

\bibitem{Contreras2017:2017JCAP...12..046C}
D.~{Contreras}, P.~{Boubel}, and D.~{Scott},
\newblock \jcap {\bf 12}, 046 (2017), 1705.06387.

\bibitem{Liu2017:2017PDU....16...22L}
G.-C. {Liu} and K.-W. {Ng},
\newblock Physics of the Dark Universe {\bf 16}, 22 (2017), 1612.02104.

\bibitem{Pospelov2009PhRvL.103e1302P}
M.~{Pospelov}, A.~{Ritz}, and C.~{Skordis},
\newblock Physical Review Letters {\bf 103}, 051302 (2009), 0808.0673.

\bibitem{Fedderke:2019ajk}
M.~A. Fedderke, P.~W. Graham, and S.~Rajendran,
\newblock (2019), 1903.02666.

\bibitem{SM}
{\em \rm See Supplemental Material at [URL will be inserted by publisher] for
  details}.

\bibitem{Caputo:2018ljp}
A.~Caputo, C.~P. Garay, and S.~J. Witte,
\newblock Phys. Rev. {\bf D98}, 083024 (2018), 1805.08780,
\newblock [Erratum: Phys. Rev.D99,no.8,089901(2019)].

\bibitem{Fujita:2018arXiv181103525F}
T.~{Fujita}, R.~{Tazaki}, and K.~{Toma},
\newblock ArXiv e-prints  (2018), 1811.03525.

\bibitem{Liu:2019brz}
T.~Liu, G.~Smoot, and Y.~Zhao,
\newblock (2019), 1901.10981.

\bibitem{Ivanov:2018byi}
M.~M. Ivanov {\em et~al.},
\newblock (2018), 1811.10997.

\bibitem{Amendola:2005ad}
L.~Amendola and R.~Barbieri,
\newblock Phys. Lett. {\bf B642}, 192 (2006), hep-ph/0509257.

\bibitem{Hlozek:2015}
R.~Hlozek, D.~Grin, D.~J.~E. Marsh, and P.~G. Ferreira,
\newblock Phys. Rev. {\bf D91}, 103512 (2015), 1410.2896.

\bibitem{Hlozek:2018}
R.~Hlozek, D.~J.~E. Marsh, and D.~Grin,
\newblock Mon. Not. Roy. Astron. Soc. {\bf 476}, 3063 (2018), 1708.05681.

\bibitem{Komatsu:2011ApJS..192...18K}
E.~{Komatsu} {\em et~al.},
\newblock \apjs {\bf 192}, 18 (2011), 1001.4538.

\bibitem{BICEP14PRL_Ade:2014xna}
P.~A.~R. Ade {\em et~al.},
\newblock Phys. Rev. Lett. {\bf 112}, 241101 (2014), 1403.3985.

\bibitem{Molinari2016PDU....14...65M}
D.~{Molinari}, A.~{Gruppuso}, and P.~{Natoli},
\newblock Physics of the Dark Universe {\bf 14}, 65 (2016), 1605.01667.

\bibitem{Planck16ParityViolation_Aghanim:2016fhp}
Planck, N.~Aghanim {\em et~al.},
\newblock Astron. Astrophys. {\bf 596}, A110 (2016), 1605.08633.

\bibitem{POLARBEAR17_Ade:2017uvt}
POLARBEAR, P.~A.~R. Ade {\em et~al.},
\newblock Astrophys. J. {\bf 848}, 121 (2017), 1705.02907.

\bibitem{SPTpol19_Wu:2019hek}
W.~L.~K. Wu {\em et~al.},
\newblock (2019), 1905.05777.

\bibitem{CMBS4:2016arXiv161002743A}
K.~N. {Abazajian} {\em et~al.},
\newblock ArXiv e-prints  (2016), 1610.02743.

\bibitem{PICO19_Hanany:2019lle}
NASA PICO, S.~Hanany {\em et~al.},
\newblock (2019), 1902.10541.

\bibitem{RPP2018Tanabashi}
Particle Data Group, M.~Tanabashi {\em et~al.},
\newblock Phys. Rev. D {\bf 98}, 030001 (2018).

\bibitem{Tseliakhovich:2010PhRvD..82h3520T}
D.~{Tseliakhovich} and C.~{Hirata},
\newblock \prd {\bf 82}, 083520 (2010), 1005.2416.

\bibitem{Conlon:2018MNRAS.473.4932C}
J.~P. {Conlon}, F.~{Day}, N.~{Jennings}, S.~{Krippendorf}, and F.~{Muia},
\newblock \mnras {\bf 473}, 4932 (2018), 1707.00176.

\bibitem{Kaufman:2016MNRAS.455.1981K}
J.~P. {Kaufman}, B.~G. {Keating}, and B.~R. {Johnson},
\newblock \mnras {\bf 455}, 1981 (2016), 1409.8242.

\bibitem{Miller:2009PhRvD..79f3008M}
N.~J. {Miller}, M.~{Shimon}, and B.~G. {Keating},
\newblock \prd {\bf 79}, 063008 (2009), 0806.3096.

\bibitem{Pagano2009PhRvD..80d3522P}
L.~{Pagano} {\em et~al.},
\newblock \prd {\bf 80}, 043522 (2009), 0905.1651.

\bibitem{Leahy:1997astro.ph..4285L}
J.~P. {Leahy},
\newblock ArXiv Astrophysics e-prints  (1997), astro-ph/9704285.

\bibitem{Alighieri:2010ApJ...715...33D}
S.~{di Serego Alighieri}, F.~{Finelli}, and M.~{Galaverni},
\newblock \apj {\bf 715}, 33 (2010), 1003.4823.

\bibitem{Whittaker:2018MNRAS.474..460W}
L.~{Whittaker}, R.~A. {Battye}, and M.~L. {Brown},
\newblock \mnras {\bf 474}, 460 (2018), 1702.01700.

\bibitem{Caputo:2019tms}
A.~Caputo {\em et~al.},
\newblock (2019), 1902.02695.

\bibitem{Alplot}
ALPlot,
\newblock {\em \rm Lichtenberg Research Group, University Mainz,},
\newblock \url{https://alplot.physik.uni-mainz.de}.

\bibitem{Ringwald:2013via}
A.~Ringwald,
\newblock {Ultralight Particle Dark Matter},
\newblock in {\em {25th Rencontres de Blois on Particle Physics and Cosmology
  Blois, France, May 26-31, 2013}}, 2013, 1310.1256.

\bibitem{Kim:1979if}
J.~E. Kim,
\newblock Phys. Rev. Lett. {\bf 43}, 103 (1979).

\bibitem{Shifman:1979if}
M.~A. Shifman, A.~I. Vainshtein, and V.~I. Zakharov,
\newblock Nucl. Phys. {\bf B166}, 493 (1980).

\bibitem{Dine:1981rt}
M.~Dine, W.~Fischler, and M.~Srednicki,
\newblock Phys. Lett. {\bf 104B}, 199 (1981).

\bibitem{Zhitnitsky:1980tq}
A.~R. Zhitnitsky,
\newblock Sov. J. Nucl. Phys. {\bf 31}, 260 (1980),
\newblock [Yad. Fiz.31,497(1980)].

\bibitem{Irsic:2017PhRvL.119c1302I}
V.~{Ir{\v s}i{\v c}}, M.~{Viel}, M.~G. {Haehnelt}, J.~S. {Bolton}, and G.~D.
  {Becker},
\newblock Physical Review Letters {\bf 119}, 031302 (2017), 1703.04683.

\bibitem{Kobayashi:2017PhRvD..96l3514K}
T.~{Kobayashi}, R.~{Murgia}, A.~{De Simone}, V.~{Ir{\v s}i{\v c}}, and
  M.~{Viel},
\newblock \prd {\bf 96}, 123514 (2017), 1708.00015.

\bibitem{Folkerts:2013tua}
S.~Folkerts, C.~Germani, and J.~Redondo,
\newblock Phys. Lett. {\bf B728}, 532 (2014), 1304.7270.

\bibitem{Kawasaki:2014una}
M.~Kawasaki, N.~Kitajima, and F.~Takahashi,
\newblock Phys. Lett. {\bf B737}, 178 (2014), 1406.0660.

\bibitem{Feix:2019lpo}
M.~Feix {\em et~al.},
\newblock JCAP {\bf 1905}, 021 (2019), 1903.06194.

\bibitem{Marsh:2018zyw}
D.~J.~E. Marsh and J.~C. Niemeyer,
\newblock (2018), 1810.08543.

\bibitem{Obata:2018vvr}
I.~Obata, T.~Fujita, and Y.~Michimura,
\newblock Phys. Rev. Lett. {\bf 121}, 161301 (2018), 1805.11753.

\bibitem{Liu:2018arXiv180901656L}
H.~{Liu}, B.~D. {Elwood}, M.~{Evans}, and J.~{Thaler},
\newblock ArXiv e-prints  (2018), 1809.01656.

\bibitem{DeRocco:2018jwe}
W.~DeRocco and A.~Hook,
\newblock Phys. Rev. {\bf D98}, 035021 (2018), 1802.07273.

\bibitem{Nagano:2019rbw}
K.~Nagano, T.~Fujita, Y.~Michimura, and I.~Obata,
\newblock (2019), 1903.02017.

\end{thebibliography}

\newpage

\appendix{

\begin{widetext}
\begin{center}
{\bf Axion-like Dark Matter Constraints from CMB Birefringence\\}
{\bf G\"unter Sigl and Pranjal Trivedi}

\end{center}

\setcounter{equation}{0}
\renewcommand{\theequation}{S\arabic{equation}}

\setcounter{page}{1}

\section{Supplementary Material}

To evaluate the behaviour of the solution to the photon equation of motion, we start with a form similar to Eq.~(A4) of \cite{Fedderke:2019ajk},
\begin{equation}
    \left( \frac{\partial^2}{\partial t^2} - \Delta \right) \vec{A} = 
    \pm i g_{a\gamma} \left[ \left( \partial_z a \right) \partial_t \vec{A} - \left( \partial_t a \right) \partial_z \vec{A} \right].
\end{equation}
Let us transform to $u-v$ coordinates: $u=z-t$ and $v=z+t$ $\Rightarrow$ $z=\frac{1}{2}(u+v)$; $z=\frac{1}{2}(v-u)$ 
\begin{equation}
    \Rightarrow \quad 
    \frac{\!\!\!\partial}{\partial z} = \frac{\!\!\!\partial}{\partial u} + \frac{\!\!\!\partial}{\partial v}\,; 
    \quad
    \frac{\!\!\partial}{\partial t} = -\frac{\!\!\!\partial}{\partial u} + \frac{\!\!\!\partial}{\partial v}\,;
\end{equation}
\begin{equation}
    \Rightarrow \quad 
    \frac{\partial a}{\partial z} \frac{\partial A}{\partial t} - 
    \frac{\partial a}{\partial t} \frac{\partial A}{\partial z} 
    = 
    \left(\frac{\partial a}{\partial v} + \frac{\partial a}{\partial u}\right) 
    \left(\frac{\partial A}{\partial v} - \frac{\partial A}{\partial u}\right) 
    -
    \left(\frac{\partial a}{\partial v} - \frac{\partial a}{\partial u}\right) 
    \left(\frac{\partial A}{\partial v} + \frac{\partial A}{\partial u}\right) 
    =
    -2 \frac{\partial a}{\partial v} \frac{\partial A}{\partial u} 
    +2 \frac{\partial a}{\partial u} \frac{\partial A}{\partial v}.
\end{equation}
The boundary condition at $v_{\rm max}$ is $\partial A / \partial v=0$; we define $\phi_a \mathrel{\mathop:}= g_{a\gamma}a / 2$
\begin{equation}\label{eq:star}
    \Rightarrow \quad 
    - \frac{\!\!\!\partial}{\partial u} \frac{\!\!\!\partial}{\partial v} A =
    \pm i \left(  \frac{\partial \phi_a}{\partial u} \frac{\partial A}{\partial v} - 
    \frac{\partial \phi_a}{\partial v} \frac{\partial A}{\partial u} \right).
\end{equation}
Now, we make the ansatz $A(u,v) = f(u) e^{\pm i \phi} + g(v) e^{\mp i \phi}$
\begin{equation}
    \Rightarrow \quad \frac{\partial A}{\partial u}  = \pm i \frac{\partial \phi}{\partial v} f(u) e^{\pm i \phi}
    + g' e^{\mp i \phi} \mp i \frac{\partial \phi}{\partial v} g(v) e^{\mp i \phi},
\end{equation}
\begin{equation}
    \Rightarrow \quad \frac{\partial^2 A}{\partial u\partial v}  = \pm i \frac{\partial^2 \phi}{\partial u\partial v} f(u) e^{\pm i \phi}
    \pm i \frac{\partial \phi}{\partial v} f' e^{\pm i \phi}
    \mp i \frac{\partial \phi}{\partial u} g' e^{\mp i \phi} 
    \mp i \frac{\partial^2 \phi}{\partial u\partial v} g(v) e^{\mp i \phi}
    -  \frac{\partial \phi}{\partial v} \frac{\partial \phi}{\partial u} A.
\end{equation}
We note that 
\begin{equation}
    \frac{\partial \phi}{\partial u} , \frac{\partial \phi}{\partial v} \sim \mathcal{O}(m_a)\phi\,; 
    \quad  \frac{\partial^2 \phi}{\partial u\partial v} \sim \mathcal{O}(m_a^2)\phi\,; \quad f', g' \sim \mathcal{O}(k)(f,g).
\end{equation}
Then the R.H.S. of Eq.~(\ref{eq:star}) is 
\begin{equation}
\pm i \left(  \frac{\partial \phi_a}{\partial u} g' e^{\mp i \phi} - 
      \frac{\partial \phi_a}{\partial v} f' e^{\pm i \phi} \right) 
    - \frac{\partial \phi_a}{\partial u} \frac{\partial \phi}{\partial v} f e^{\pm i \phi}
    + \frac{\partial \phi_a}{\partial u} \frac{\partial \phi}{\partial v} g e^{\mp i \phi}
    + \frac{\partial \phi_a}{\partial v} \frac{\partial \phi}{\partial u} f e^{\pm i \phi}
    - \frac{\partial \phi_a}{\partial v} \frac{\partial \phi}{\partial u} g e^{\mp i \phi}
\end{equation}
which implies that the non-linear terms do not cancel but are suppressed by $(m_a/k)$ relative to the other terms.\\
\\
If $m_a \ll k$, then to first order in $\phi$ the L.H.S. of Eq.~(\ref{eq:star}) is 
\begin{equation}
\pm i \frac{\partial \phi_a}{\partial u} g' e^{\mp i \phi} 
\mp i \frac{\partial \phi_a}{\partial v} f' e^{\pm i \phi},
\end{equation}
to first order in $\phi$, $\phi_a$ one thus has,
\begin{equation}
      \frac{\partial^2 \phi}{\partial u\partial v} f(u) 
    + \frac{\partial \phi}{\partial v} f'(u)
    = \frac{\partial \phi_a}{\partial v} f'(u),
\end{equation}
\begin{equation}
      \frac{\partial^2 \phi}{\partial u\partial v} g(v) 
    + \frac{\partial \phi}{\partial u} g'(v)
    = \frac{\partial \phi_a}{\partial u} g'(v).
\end{equation}
For $f(u) \propto e^{i k u}$ one has $f'(u)=ikf(u)$,
\begin{equation}
     \Rightarrow \quad 
     \frac{\partial^2 \phi}{\partial u\partial v} 
    + ik \left( \frac{\partial \phi}{\partial v}
    - \frac{\partial \phi_a}{\partial v} \right) = 0,
\end{equation}
whose general solution is
\begin{equation}\label{eq:phi_sol}
    \phi(u,v) = ik\int^u\phi_a(u',v) e^{ik\left( u'-u \right)} du' +c(v)e^{-iku}.
\end{equation}
Further, because 
\begin{equation}
    \frac{\partial \phi}{\partial u} = ik \phi_a(u,v) - ik \phi(u,v),
\end{equation}
\begin{equation}
     \Rightarrow \quad 
     \frac{\partial^2 \phi}{\partial v\partial u} 
    = ik \left( \frac{\partial \phi_a}{\partial v}
    - \frac{\partial \phi}{\partial v} \right).
\end{equation}
The boundary conditions, in general, eliminate the homogeneous term $\propto c(v)$. If $\phi_a$ varies very slowly, it can be pulled out in front of the integral in Eq.~(\ref{eq:phi_sol}) so that, 
\begin{equation}
    \phi(u,v) \sim \phi_a(u,v)
\end{equation}
for the case $m_a \ll k$. This is equivalent to neglecting second derivatives of the photon phase $\phi$. 

We conclude that for CMB birefringence due to ALPs, where $g_{a\gamma}$ constraints can be improved over the range $10^{-27} \lesssim m_a \lesssim 10^{-24}$ eV, the fact that $m_a \ll k$ implies that non-linear terms are highly suppressed and we can use $\phi \sim g_{a\gamma} a /2$ or Eq.\,(4), $\Delta \phi \sim g_{a\gamma} \Delta a$. 
The ratio $(m_a/k) \sim 1$ for much larger ALP masses $m_a \sim 10^{-6}$ eV where non-linear terms won't be negligible but then the $g_{a\gamma}$ constraints will be far too weak to be of interest.

 \end{widetext}

}

\end{document}